# Solar Flare Index Prediction Using SDO/HMI Vector Magnetic Data Products with Statistical and Machine-learning Methods

Hewei Zhang[1,2], Qin Li[2,3], Yanxing Yang[4], Ju Jing[2,3], Jason T. L. Wang[2,5], Haimin Wang[2,3], and Zuofeng Shang[1,2]
[1] Department of Mathematical Sciences, New Jersey Institute of Technology, Newark, NJ 07102-1982, USA
[2] Institute for Space Weather Sciences, New Jersey Institute of Technology, Newark, NJ 07102-1982, USA
[3] Big Bear Solar Observatory, New Jersey Institute of Technology, 40386 North Shore Lane, Big Bear, CA 92314, USA
[4] Department of Physics, New Jersey Institute of Technology, Newark, NJ 07102-1982, USA
[5] Department of Computer Science, New Jersey Institute of Technology, Newark, NJ 07102-1982, USA

## Abstract

Solar flares, especially the M- and X-class flares, are often associated with coronal mass ejections. They are the most important sources of space weather effects, which can severely impact the near-Earth environment. Thus it is essential to forecast flares (especially the M- and X-class ones) to mitigate their destructive and hazardous consequences. Here, we introduce several statistical and machine-learning approaches to the prediction of an active region's (AR) flare index (FI) that quantifies the flare productivity of an AR by taking into account the number of different class flares within a certain time interval. Specifically, our sample includes 563 ARs that appeared on the solar disk from 2010 May to 2017 December. The 25 magnetic parameters, provided by the Space-weather HMI Active Region Patches (SHARP) from the Helioseismic and Magnetic Imager on board the Solar Dynamics Observatory, characterize coronal magnetic energy stored in ARs by proxy and are used as the predictors. We investigate the relationship between these SHARP parameters and the FI of ARs with a machine-learning algorithm (spline regression) and the resampling method (Synthetic Minority Oversampling Technique for Regression with Gaussian Noise). Based on the established relationship, we are able to predict the value of FIs for a given AR within the next 1 day period. Compared with other four popular machine-learning algorithms, our methods improve the accuracy of FI prediction, especially for a large FI. In addition, we sort the importance of SHARP parameters by the Borda count method calculated from the ranks that are rendered by nine different machine-learning methods.

*Unified Astronomy Thesaurus concepts:* Solar flares (1496); Solar physics (1476); Solar activity (1475)

## 1. Introduction

Solar flares are enhanced emissions across the electromagnetic spectrum that occur on a minute-to-hour time frame (Benz 2017). The frequency of solar flares varies with the solar cycle, accounting for an increasing frequency every 11 yr. Small flares usually only arouse an increase in intensities of visible light, soft X-ray, and radio waves, releasing energy in a magnitude of $10^{28} \sim 10^{29}$ erg. However, a major (M-class or X-class) flare eruption can release energy as much as $4 \times 10^{32}$ erg, of which electromagnetic radiation only accounts for one-quarter. The rest of the energy is released by high-energy particles in the form of a plasma cloud moving at a speed of 1500 km s$^{-1}$. In particular, solar flares and the often associated coronal mass ejections are the drivers of "space weather." Geomagnetic storms caused by them can disrupt or damage spacecraft, harm astronauts and high-altitude pilots, interrupt communications and navigation systems, and even shut down portions of the electric transmission system when they reach Earth (Schwenn 2006). Therefore, the reliable prediction of flares, especially large flares, is a very urgent task.

A solar flare usually originates from an active region (AR), which is an area with enhanced magnetic fields on the Sun. Although the triggering mechanism of flares still remains elusive, it is certain that the free magnetic energy available to power a flare is stored in the host AR. Generally speaking, the more the AR's magnetic field deviates from a simple potential configuration, the more free magnetic energy is stored in the AR and the more likely it is that a flare will occur in the AR. However, the coronal magnetic fields cannot be precisely measured for the time being and, most likely, in the foreseeable future. Therefore the flare forecasting efforts have been made almost exclusively with magnetic parameters that are derived from the photospheric magnetic fields (Toriumi & Wang 2019).

In the past few decades, traditional statistical methods have been extensively used to predict solar flares. Many statistical studies have focused on the distribution of solar flare intervals (Pearce et al. 1993; Boffetta et al. 1999; Lepreti et al. 2001; Kubo 2008; Wheatland 2010). Some statistical models are based on the morphological characteristics of flare (Gallagher et al. 2002; Stepanov et al. 2004; Wheatland 2004, 2005; Zharkov & Zharkova 2006; Barnes et al. 2007; Song et al. 2009; Yu et al. 2009; Bloomfield et al. 2012; McCloskey et al. 2016; Leka et al. 2018). For example, Gallagher et al. (2002) and Bloomfield et al. (2012) adopt the Poisson model to estimate the probability of solar flares. Song et al. (2009) predict the probability of flares in each AR during the next 24 hr by using the ordinal logistic regression method. However, the prediction accuracy of these traditional statistical methods for flares, especially large flares, is not satisfactory.

With the explosion of digital data, the human processing ability is being challenged. Machine-learning methods, as a state-of-the-art tool, have attracted more and more attention. Several machine-learning algorithms have been applied to the flare prediction problem: support vector machines (Li et al. 2007; Qahwaji & Colak 2007; Yuan et al. 2010; Bobra & Couvidat 2015; Muranushi et al. 2015; Nishizuka et al. 2017), random forest (RF; Liu et al. 2017; Florios et al. 2018; Wang et al. 2019; Hazra et al. 2020), k-nearest neighbors (Li et al. 2008; Huang et al. 2013; Winter & Balasubramaniam 2015; Hazra et al. 2020), LASSO (Benvenuto et al. 2018; Jonas et al. 2018), decision trees (Yu et al. 2009, 2010), LSTM (Chen et al. 2019; Liu et al. 2019a; Jiao et al. 2020; Wang et al. 2020; Sun et al. 2022), and CNN (Huang et al. 2018; Park et al. 2018; Zheng et al. 2019; Zhu et al. 2020; Sun et al. 2022). These studies contributed significantly to the forecasting of solar eruptions, which still needs to be improved methodologically due to the shortcomings regarding the sampling. The main problem in this field is that the data sample is imbalanced, which leads to a poor prediction of large flares. Shin et al. (2016) adopt multiple linear regression and artificial neural network methods to forecast the maximum flare flux for strong flares. Kusano et al. (2020) present a physics-based model to predict large solar flares through a critical condition of magnetohydrodynamic instability, triggered by magnetic reconnection. In this paper, we develop the spline regression model (De Boor & Höllig 1982) to forecast the quantitative flare, and then compare and analyze prediction performance with other popular prediction models, i.e., linear regression, RF, LASSO and Gaussian process regression (GPR). In addition, to solve the problem of data imbalance, we adopt an advanced resampling method, the Synthetic Minority Oversampling Technique for Regression with Gaussian Noise (SMOGN; Branco et al. 2017), to improve the prediction accuracy of large flares.

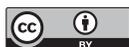

**Table 1**
The Number of Flare Indices of Four Groups from 2010 May 1 to 2017 December 28 in the Sample

| Group/Year | 2010 | 2011 | 2012 | 2013 | 2014 | 2015 | 2016 | 2017 | Total |
|---|---|---|---|---|---|---|---|---|---|
| FI < 1 | 91 | 100 | 85 | 91 | 22 | 74 | 144 | 77 | 684 |
| 1 <= FI < 10 | 51 | 202 | 234 | 262 | 231 | 214 | 93 | 60 | 1347 |
| 10 <= FI < 100 | 5 | 60 | 71 | 82 | 95 | 79 | 12 | 12 | 416 |
| FI <= 100 | 0 | 8 | 11 | 6 | 19 | 7 | 0 | 6 | 57 |
| Total | 147 | 370 | 401 | 441 | 367 | 374 | 249 | 255 | 2504 |

Since 2010 May, the Helioseismic and Magnetic Imager (HMI; Schou et al. 2012) on board the Solar Dynamics Observatory (SDO; Pesnell et al. 2011) has been continuously producing full-disk photospheric vector magnetograms at a 12 minute cadence. In particular, the Space-weather HMI Active Region Patches (SHARPs; Bobra et al. 2014; Bobra & Couvidat 2015) released by the SDO-HMI team include automatically identified and tracked ARs in map patches and contain several essential magnetic parameters. Since then, these SHARP parameters, as the predictors, have been widely used in many flare prediction studies (Liu et al. 2017, 2019a; Jonas et al. 2018; Chen et al. 2019; Wang et al. 2019, 2020; Jiao et al. 2020; Sun et al. 2022).

The goal of our research is to predict the flare productivity of a given AR, quantified by flare indexes (FIs; Jing et al. 2006; Song et al. 2009; Jiao et al. 2020), in the next 1 day period using 25 SHARPs parameters. To this end, we explore the relationship between FI and SHARP parameters as follows. First, we apply a popular family of power transformations for achieving approximate normality, i.e., Yeo–Johnson power transformations (Yeo & Johnson 2000). Then we test statistical significance and linearity for each SHARP parameter, and perform an exhaustive sieving method to exclude some highly correlated features from our model. Next, based on the feature selection result, we apply the spline regression technique to establish the relationship between the selected SHARP parameters and FI, from which we are able to predict the value of FI for a given AR. To address the data imbalance problem, we adopt an advanced resampling method, SMOGN, to improve the prediction accuracy for a large FI. Finally, we compare the performance of four popular machine-learning methods (i.e., linear regression, LASSO, RF, and GPR) with our method. In addition, to better understand the contribution of these SHARP parameters on the FI prediction, we sort the importance of SHARP parameters on FI prediction by the Borda count score calculated from the ranks that are rendered by nine different machine-learning methods.

The paper is organized in the following manner. Section 2 includes data preparation (Section 2.1) and feature selection (Section 2.3) for regression tasks. Section 3 introduces the prediction algorithm (Section 3.1) and the solution (Section 3.2) to the problem of data imbalance. Section 4 reports magnetic parameter ranking results and the corresponding physical explanation. The conclusions and future works are presented in the Section 5.

## 2. Data Preparation

### 2.1. Details of Data

The overall flare productivity of a given AR has been quantified by the soft X-ray (SXR) flare index (FI) (Antalova 1996; Abramenko 2005; Jing et al. 2006, 2010; Song et al. 2009). Specifically, flares, from weak to strong, are classified as B, C, M, or X according to their peak SXR flux (of $10^{-7}$, $10^{-6}$, $10^{-5}$, and $10^{-4}$ W m$^2$ magnitude order, respectively), as measured by the Geostationary Operational Environmental Satellite (GOES). FI is calculated by weighting the GOES SXR flares classes of B, C, M, and X as 0.1, 1, 10, and 100, respectively, within a certain time window $\tau$, i.e.,

$$\text{FI} = 0.1 \times \sum_\tau I_B + 1 \times \sum_\tau I_C + 10 \times \sum_\tau I_M + 100 \times \sum_\tau I_X,$$

where $I_B$, $I_C$, $I_M$, and $I_X$ are GOES peak intensities of B-, C-, M-, and X-class flares produced by the given AR over the period $\tau$, and $\tau$ is selected to be 1 day in this study to account for the flare production generated from an AR on the solar disk during a day starting from 0:00 Universal Time (UT).

Our sample includes 563 ARs from 2010 May to 2017 December. For each AR, we calculate the FI of these ARs for each day during their disk passage. Thus a total of 2504 FIs of 563 ARs are acquired. As shown in Table 1, the data are categorized as four groups according to the magnitude of the FI, i.e., FI < 1; 1 ⩽ FI < 10; 10 ⩽ FI < 100; FI ⩾ 100, equivalent to a daily average of a B-class or less, a C-class, a

**Table 2**
25 SDO/HMI Magnetic Parameters

| Keyword | Description |
|---|---|
| TOTUSJH | Total unsigned current helicity |
| TOTBSQ | Total magnitude of Lorentz force |
| TOTPOT | Total photospheric magnetic free energy density |
| TOTUSJZ | Total unsigned vertical current |
| ABSNJZH | Absolute value of the net current helicity |
| SAVNCPP | Sum of the modulus of the net current per polarity |
| USFLUX | Total unsigned flux |
| AREA_ACR | Area of strong field pixels in the active region |
| MEANPOT | Mean photospheric magnetic free energy |
| R_VALUE | Sum of flux near polarity inversion line |
| SHRGT45 | Fraction of area with shear $>45°$ |
| MEANSHR | Mean shear angle |
| MEANGAM | Mean angle of field from radial |
| MEANGBT | Mean gradient of total field |
| MEANGBZ | Mean gradient of vertical field |
| MEANGBH | Mean gradient of horizontal field |
| MEANJZH | Mean current helicity |
| MEANJZD | Mean vertical current density |
| MEANALP | Mean characteristic twist parameter, $\alpha$ |
| TOTFX | Sum of $x$-component of Lorentz force |
| TOTFY | Sum of $y$-component of Lorentz force |
| TOTFZ | Sum of $z$-component of Lorentz force |
| EPSX | Sum of $x$-component of normalized Lorentz force |
| EPSY | Sum of $y$-component of normalized Lorentz force |
| EPSZ | Sum of $z$-component of normalized Lorentz force |

M-class, and a X-class flare, respectively. Forecasting strong flares (M- or X-class) is particularly important because of their space weather effects. In our data set, 19% FIs are greater than 10, and only 2% FIs are greater than 100. In this work, we emphasize to improving the predictive accuracy of FI greater than 10.

The SDO/HMI team has been releasing SHARPs since 2012 (Bobra et al. 2014), which can be found at the Joint Science Operations Center website. The 25 SHARP parameters of ARs are used as the FI predictors in this work, and are listed in Table 2. These SHARP parameters characterize physical properties of the AR, and are generally classified as intensive (spatial averages, e.g., MEANPOT), or extensive (summations or integrations, e.g., TOTPOT) measures (Welsch et al. 2009).

### 2.2. Data Transformation

The classic spline regression technique requires the model error being Gaussian (Wahba 1990). The genuine data on solar flares, on the other hand, shows a lot of non-Gaussianity. As a result, data must be preprocessed before using the spline regression approach. In this paper, we apply the power transformation method (Box & Cox 1964), which is widely used in the statistical literature. SHARP parameters do not exhibit any explicit FI dependence in the original data—see Figure 1 (blue), whereas certain correlations between them emerge after the processing by power transformation—see Figure 1 (red), providing the possibility to predict solar flares by SHARP parameters.

One-parameter Box–Cox (Box & Cox 1964) and two-parameter Yeo–Johnson transformations (Yeo & Johnson 2000) are the two branches that are widely used in the implementation of power transformation. In particular, the Box–Cox family includes two different types of transformation based on the selection of $\lambda$:

$$y^{(\lambda)} = \begin{cases} (y^\lambda - 1)/\lambda & \text{if } \lambda \neq 0 \\ \log(y) & \text{if } \lambda = 0 \end{cases}, \quad (1)$$

where $y$ is a set of strictly positive numbers and $\lambda$ the power parameter. The logarithmic form in Equation (1) is a special case only when the $\lambda$ is selected to be zero. Hence, the existence of nonpositive numbers in the SHARP parameters are forbidden as the values of variable, making the Box–Cox transformation not applicable in this study. The Yeo–Johnson transformation was proposed in 2000, preserving many good features of the Box–Cox power family without the restriction that the values of $y$ must be strictly positive:

$$\psi(\lambda, y) = \begin{cases} ((y+1)^\lambda - 1)/\lambda & \text{if } \lambda \neq 0, y \geq 0 \\ \log(y+1) & \text{if } \lambda = 0, y \geq 0 \\ -[(-y+1)^{2-\lambda} - 1]/(2-\lambda) & \text{if } \lambda \neq 2, y < 0 \\ -\log(-y+1) & \text{if } \lambda = 2, y < 0 \end{cases}, \quad (2)$$

wherein the transformation has the same form as that of the Box–Cox with only the replacing of $y$ by $y + 1$ for a strictly positive $y$, and the replacing of $y$ and $\lambda$ by $-y + 1$ and $2 - \lambda$ for a strictly negative $y$, respectively.

### 2.3. Feature Selection

We describe our procedure for selecting the SHARP parameters. Our approach is based on the intuition that significant SHARP parameters should have strong marginal association with the observed flare index, for which a marginal nonparametric regression is fitted between the flare index and each individual SHARP parameter. Based on this, we coarsely select 18 SHARP parameters. Moreover, we propose a new method called an *exhaustive sieve* to refine the result.

#### 2.3.1. Coarse Screening

First, a nonparametric model for the FI and SHARP parameters is developed:

$$y = f_j(X_j) + \sigma\varepsilon, \quad j = 1,\ldots,25, \quad (3)$$

where $y$ is the response variable, $X_1,\ldots,X_{25} \in [0, 1]$ are feature variables, $\epsilon$ is zero-mean continuous random variables with finite standard deviation $\sigma$, and $f_j$ is an unknown function belonging to an $m$-order periodic Sobolev space $S^m$ on [0, 1], which characterizes the marginal association between $y$ and $X_j$. One can regard $y$ and $X_j$ as the observed value of FIs and the 25 transformed SHARP parameters, and so, $f_j$ represents their functional relationship. We are interested in making valid statistical inferences regarding $f_i$ using techniques such as estimation or hypothesis testing.

Here, we propose a nonparametric hypothesis testing approach for testing the significance of each feature $X_1$, $X_2$, ..., $X_{25}$ and exploring the correlation between the response variable and the feature variables. Our approach is motivated from the nonparametric inferential literature (Shang 2010; Shang & Cheng 2013, 2015, 2017; Cheng & Shang 2015; Liu et al. 2019b, 2020, 2021, 2022; Yang et al. 2020). In this section, we test the marginal effect of each variable. Correlation analysis will be conducted in Section 2.3.2.

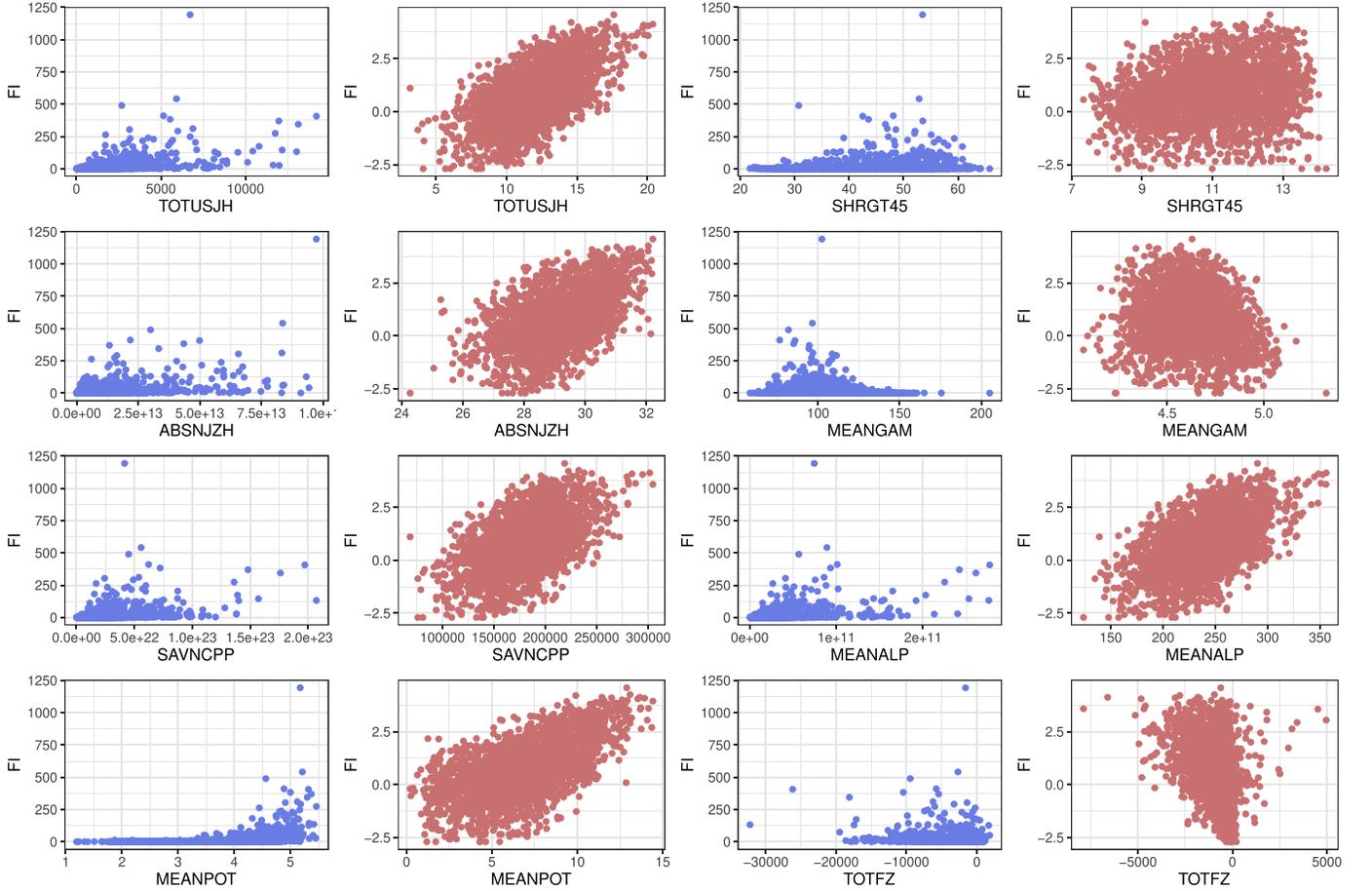

**Figure 1.** Evolution of FI with respect to eight SHARP parameters, i.e., TOTUSJH, ABSNJZH, SAVNCPP, MEANPOT (first and second columns) and SHRGT45, MEANGAM, MEANALP, TOTFZ (third and fourth columns). The effect of power transformation is characterized by plotting the original data (blue) as well as the corresponding transformed data (red) in the same row.

The hypothesis for testing the significance of each individual SHARP parameter is

$$H_0: f_j \text{ is constant versus } H_1: f_j \text{ is nonconstant.} \quad (4)$$

Specifically, our test statistics for Equation (4) is $T = \|\widehat{f}_j - \bar{y}\|_{L^2}^2$ in which $\widehat{f}_j$ is a smoothing spline estimator of $f_j$ based on Equation (3) and $\bar{y}$ is the averaged FIs. We reject $H_0$ at a 1% significance level if $T > 2.576\sigma^2$ with $\sigma^2$ the theoretical variance of $T$ under $H_0$ (see Shang & Cheng 2017; Yang et al. 2020). Rejection of $H_0$ implies that $X_j$ contributes significantly to the model. If $H_0$: $f_j$ = constant is not rejected, then this indicates that $X_j$ can be removed from the model.

Including an irrelevant variable in the model might actually raise the mean square error, reducing the model's effectiveness. This test assists in determining the value of each regression variable contained inside the regression model. Seven SHARP parameters failed to present significance in our study. The p-values in Figure 2 reveal that the p-values for these seven parameters (MEANSHR, MEANGBT, MEANGBZ, MEANGBH, MEANJZD, EPSY and EPSZ) are greater than 0.01, indicating that these parameters can be removed due to lack of statistical significance.

The second step is to determine whether the relationship between the FI and the remaining SHARP variables is linear. Here, we apply the *correlation coefficient* (Pearson 1896) to assess a possible linear association between the FI and each SHARPs. Correlation coefficients (Equation (5)) range from −1 to +1, with ±1 denoting an ideal positive/negative linear relationship and 0 denoting the absence of a linear relationship:

$$r = \frac{\sum_{i=1}^{n}(x_i - \bar{x})(y_i - \bar{y})}{\sqrt{\left[\sum_{i=1}^{n}(x_i - \bar{x})^2\right]\left[\sum_{i=1}^{n}(y_i - \bar{y})^2\right]}}. \quad (5)$$

A condition that is necessary for a perfect correlation is that the shapes of the individual X data and the individual Y data must be identical (Ratner 2009). Additionally, correlations must be shown to be statistically significant before the correlation coefficient is considered meaningful (Taylor 1990). Thus, power transformation and significance testing provide a robust theoretical foundation for correlation analysis in our work. The correlation coefficients between SHARPs and the FI with a cutoff of 0.5 are shown in Figure 2 under the assumption of a large sample size ($n > 100$, see Taylor 1990). Nine SHARP parameters (TOTUSJH, TOTBSQ, TOTPOT, TOTUSJZ, ABSNJZH, SAVNCPP, USFLUX, MEANPOT, MEANALP) represent large correlation coefficients ($r > 0.5$), indicating that the linear relationships between these nine SHARP variables and the FI are statistically significant.

Therefore, based on the above significance test and linear correlation analysis, we establish the following preliminary

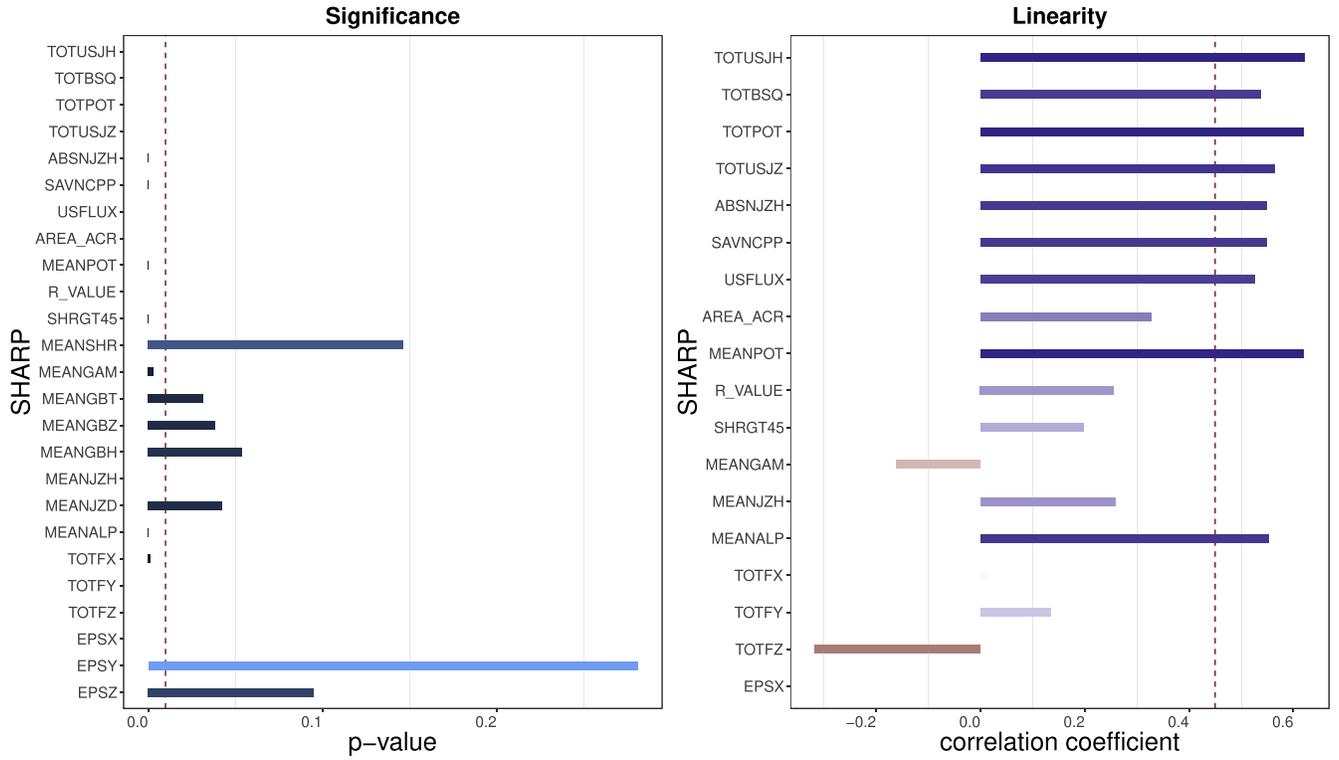

**Figure 2.** The histogram on the left shows *p*-values of 25 SHARP parameters with respect to the significance test. The histogram on the right gives the correlation coefficient value of the rest 18 SHARP variables. The vertical dashed lines highlight critical values used to define parameters that are statistically significant (*p*-value = 0.01 in left panel) and linearly correlated (*r* = 0.5 in right panel).

model:

$$\begin{aligned} \text{FI} &= \beta_1 * \text{TOTUSJH} + \beta_2 * \text{TOTBSQ} + \beta_3 * \text{TOTPOT} \\ &+ \beta_4 * \text{TOTUSJZ} \\ &+ \beta_5 * \text{ABSNJZH} + \beta_6 * \text{SAVNCPP} \\ &+ \beta_7 * \text{USFLUX} + \beta_8 * \text{MEANPOT} + \beta_9 * \text{MEANALP} \\ &+ f_1(\text{AREA\_ACR}) \\ &+ f_2(\text{R\_VALUE}) + f_3(\text{SHRGT45}) \\ &+ f_4(\text{MEANGAM}) + f_5(\text{MEANJZH}) + f_6(\text{TOTFX}) \\ &+ f_7(\text{TOTFY}) + f_8(\text{TOTFZ}) \\ &+ f_9(\text{EPSX}) + \varepsilon_i, \quad i = 1, \ldots, 18, \end{aligned}$$

(6)

where $\beta_i$, $i = 1,\ldots,9$ are the coefficients for the nine linear SHARP parameters. And $f_i$, $i = 1,\ldots,9$, are the functions of the nine nonlinear SHARP parameters.

*2.3.2. Exhaustive Sieve*

While we exclude seven SHARP parameters based on their importance in the hypothesis test, the resulting model remains relatively complicated and may be impacted by the high correlation between several components. Figure 3 gives the symmetric correlation matrix for the 25 SHARP parameters with color coding. A correlation matrix is a table that displays the correlation coefficients between sets of variables. Each cell in the table represents the correlation between two SHARP parameters. Green and yellow represent positive and negative correlation coefficient values, respectively. The darker the color, the closer the absolute value of the correlation coefficient is to 1, indicating a stronger linear relationship between variables. We recognize that some SHARP parameters are substantially correlated based on this correlation matrix. Notably, excessive correlation can diminish the precision of estimated coefficients, leading to skewed or misleading findings, which weakens the statistical power of the regression model. To address this problem, we perform an exhaustive sieving method to exclude some highly correlated features. The three steps are described as follows.

1. We begin by grouping the highly correlated features according to their correlation coefficients. Group 1: TOTUSJH, TOTBSQ, TOTPOT, MEANALP, SAVNCPP, USFLUX; Group 2: MEANGAM, MEANGBT; Group 3: R_VALUE, MEANJZH, SHRGT45; Group 4: TOTUSJZ, ABSNJZH; Group 5: MEANSHR, MEANJZD
2. We eliminate one or more parameters from each group using the exhaustive sieve approach and compare the performance of each combination of various groups.
3. Finally, we obtain the optimal combination and construct the model.

The exhaustive sieved model is as follows:

$$\begin{aligned} \text{FI} &= \beta_1 * \text{TOTUSJH} + \beta_2 * \text{MEANALP} + \beta_3 * \text{USFLUX} \\ &+ \beta_4 * \text{TOTPOT} \\ &+ \beta_5 * \text{SAVNCPP} + \beta_6 * \text{MEANPOT} \\ &+ f_1(\text{SHRGT45}) + f_2(\text{MEANGAM}) + f_3(\text{AREA\_ACR}) \\ &+ f_4(\text{EPSX}) + f_5(\text{TOTFZ}) + \varepsilon_i, \quad i = 1,\ldots,11. \end{aligned}$$

(7)

## 3. Implementation of Flare Index Prediction

*3.1. Multivariate Spline Regression*

We develop the FI prediction model using the multivariate spline regression technique in accordance with the exhaustive sieved model above. Spline regression is one of the most important nonlinear regression techniques. Rather than creating a single model for the entire data set, spline regression divides it into $k$ continuous bins via knots and fits each bin with a separate model, such as a linear function or low-order polynomial function (such as quadratic or cubic multinomial, etc.). Splines are continuous, piecewise polynomial functions. To fit a piecewise function, it is self-evident that the more knots in the model, the more flexible it is. While spline regression can be considered piecewise regression, it is not straightforward piecewise regression; rather, it is piecewise regression with constraints, which needs continuity at each knot. In this study, we perform regression using natural cubic spline.

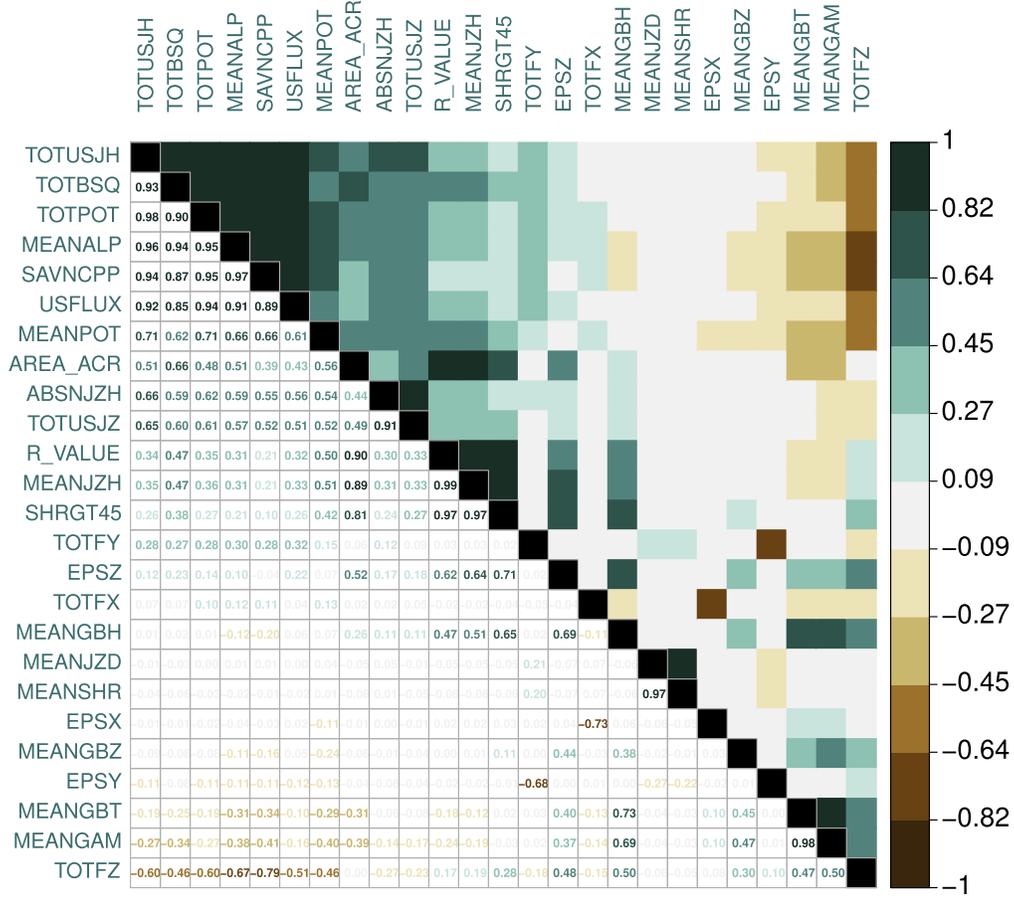

**Figure 3.** Correlation matrix that displays correlation coefficients for 25 SHARP parameters. The upper and lower triangles are correspondingly filled with colors and numeric values. Positive correlations are displayed in green and negative correlations in yellow. Color intensity of the cell is proportional to the correlation coefficients.

*B*-spline or basis spline is a commonly used spline basis of spline functions: any spline function of a given degree can be represented as a linear combination of *B*-splines of the same degree. A *B*-spline curve is defined as the following:

$$f(t) = \sum_{i=0}^{n} B_{i,d}(t) C_i, \qquad (8)$$

where $B_{i,d}(t)$ represents the *B*-spline basis functions at the scalar $t_i$ with the degree of $d$. Note that, the sequence of $t_i$ is referred to as nondecreasing knots where $0 \leqslant i \leqslant n + d + 1$. $C_i$ denotes control points for which the collection of $n + 1$ control points may be considered column vectors $\hat{C}$:

$$\hat{C} = (C_0, C_1, \ldots, C_n)^\top. \qquad (9)$$

Consider $\hat{S}$ to be the collection of the sample data $S_p$, where

$$\hat{S} = (S_0, S_1, \ldots, S_p)^\top. \qquad (10)$$

The least-squares error function between the *B*-spline curve and the sample points can then be represented as

$$E(\hat{C}) = \frac{1}{2} \sum_{p=0}^{m} \left| \sum_{j=0}^{n} B_{j,d}(t_p) C_j - S_p \right|^2. \qquad (11)$$

By selecting the control points, this error function may be minimized to the greatest extent possible.

*B*-splines can be used to represent any spline (of any degree). Degree = 3 is a common choice. The spline is referred to as a cubic spline in this situation. Typically, the second derivative of each third-order polynomial is set to zero at the endpoints in order to assure smoothness at the data points. This is known as natural cubic spline, which has a lower tendency for data points to fluctuate.

In our model, we employ both cubic and natural cubic *B*-splines. The tool we utilized here is *bs* and *ns* from the *splines* R package. We collect data samples from 2010 to 2016 as training sets, and those in 2017 as testing sets. Given the comparatively large FI in 2017, utilizing 2017 data to assess our model's predictive power for a large FI is an effective way to validate our model's predictive power for a large FI. The number of knots associated with each parameter is determined by the predicted RMSE result, that is, the number of knots that

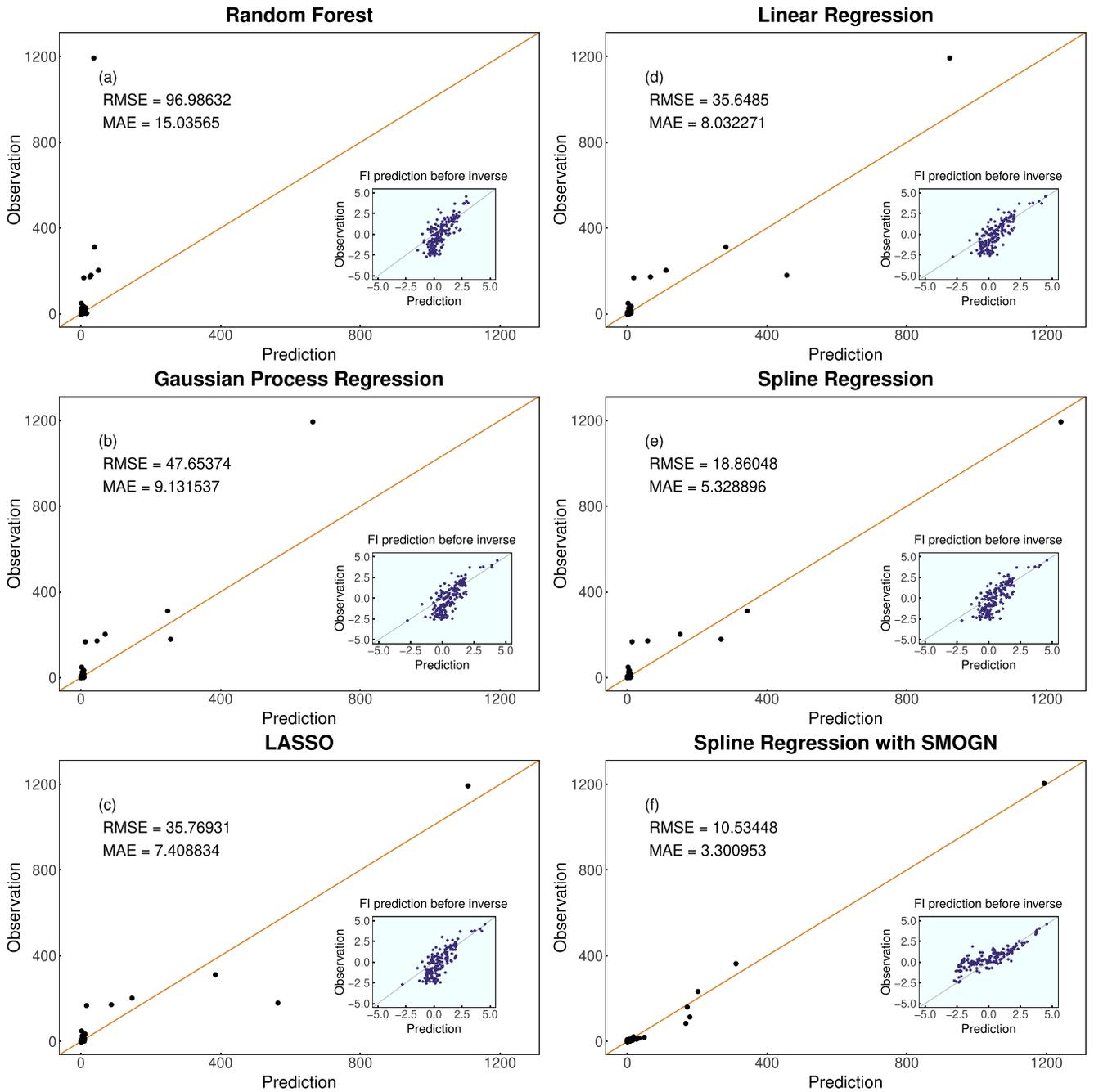

**Figure 4.** Performance of FI prediction using different models, i.e., random forest (a), LASSO (b), Gaussian process regression (c), linear regression (d), spline regression (e), and spline regression with SMOGN (f), is estimated by plotting dots at coordinates (predicted FI, observed FI). The reference line (yellow) highlights the ideal case wherein the prediction is strictly the same as the observation. Inner panels show the predicted FI without inverse processing vs. observed FI by power transformation.

minimizes the obtained RMSE. Figure 4 compares the model performance of spline regression to that of four popular machine-learning methods (i.e., linear regression, LASSO, RF, and GPR). Since we preprocess the data by power transformation, we should inverse the results after the prediction is complete. The outer and inner panels, respectively, display the prediction results before and after the inverse transformation. Due to the fact that R squared is not appropriate to compare the linear model and nonlinear model (Kvålseth 1985), RMSE (Equation (12)) and mean absolute error (MAE; Equation (13)) are selected as metrics for evaluating the predictive effect of the FI. As illustrated in Figure 4, the RMSE and MAE of spline regression both exhibit the lowest values among the five tested algorithms, illustrating the advantage of our model. The graph in the lower-right corner of Figure 4 is a combination of spline regression and SMOGN, which will be discussed in the following section. Here, we implement RF using the randomForest function (Liaw & Wiener et al. 2002) in R package randomForest. Furthermore, we fine-tune the model by defining a grid of algorithm with the following functions: trainControl from the caret package to conduct repeated 5-fold cross-validation; and expand.grid from the base package to get all combinations of hyperparameters, i.e., *mtry* is selected between 1 and 10, *maxnode* and *nodesize* are

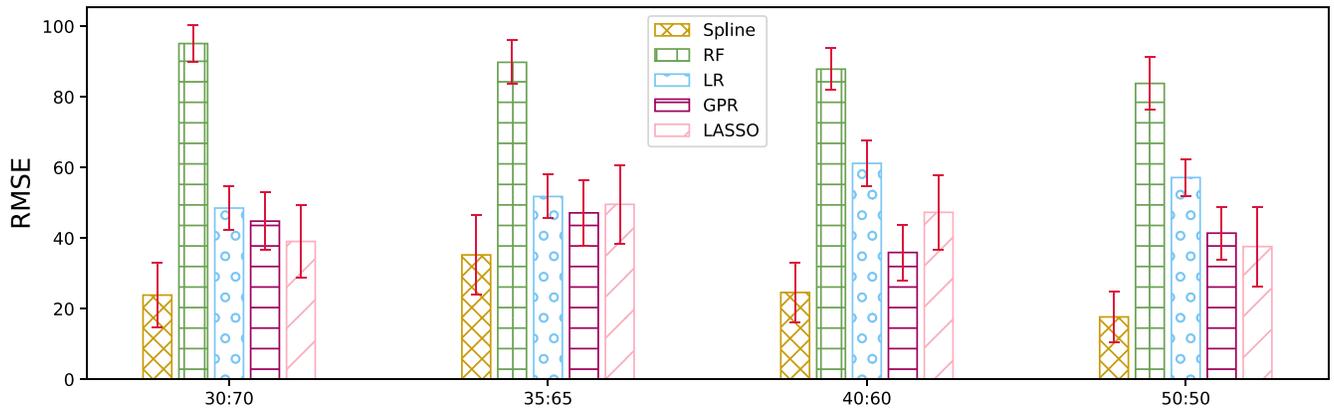

**Figure 5.** RMSE of predicted FI results from resampled data sets by SMOGN algorithm with fixed sampling ratios of minority to majority (i.e., 30:70, 35:75, 40:60, and 50:50), respectively. Standard deviations of RMSE are represented by red error bars.

selected between 2 and 20, and *ntree* is selected from 100, 250, 300, 350, 400, 450, 500, 550, 600, 800, 1000, 2000, 3000. The following parameters are included in the final optimized model: *mtry* = 3, *maxnode* = 17, *nodesize* = 6, and *ntree* = 100. We construct GPR using the gausspr function (Williams & Barber 1998) in the kernlab R package with type = "regression" and kernel = "polydot." The kernel parameter is selected with 5-fold cross-validation from the eight most common kernel functions in the package; For LASSO, we conduct 5-fold cross-validation using the cv.glmnet R function (Simon et al. 2011) and get the lasso tuning parameter = lambda.min which minimizes the cross-validation error. Noting that RF is not robust to single outliers, the RMSE of RF is high due to the existence of a large FI in 2017. However, it outperformed models other than spline regression when test data sets are more balanced. Therefore, we will continue to examine this model as we add additional data in the future.

$$\text{RMSE} = \sqrt{\frac{1}{n}\sum_{i=1}^{n}(Y_i - \hat{Y}_i)^2} \quad (12)$$

$$\text{MAE} = \frac{1}{n}\sum_{i=1}^{n}|Y_i - \hat{Y}_i|. \quad (13)$$

In the knots selection process of spline regression, if the training set and test set are quite specialized, it is easy to run into overfitting difficulties. To verify the validity of our model, we generate up to 500 different training and testing folds of our data sample. To be specific, 90% of data samples from 2010 to 2016 were randomly selected as the training set, while 90% of the data samples from 2017 were randomly selected as the test set. Then fold them together. Thus, each fold represents a unique combination of training and test data that can be used to estimate the performance of our algorithm. The averages of RMSE and MAE are 18.9858 and 4.1932, respectively.

The prediction of strong flares is of significance in mitigating space weather effects, whereas weak flares allow us to assess the reliability of the methodology in this study. Thus, we conduct an additional seven predictions using the data from the year 2010 to 2016 as the test sets, respectively. Each of these seven predictions is produced by training the models using the data from the rest years. The aggregated performance is represented as boxplots—see Figures 6(a)–(c). The spline regression model exhibits the best performance in the predictions of strong, weak, and all FIs.

### 3.2. Data Imbalance Problem

In almost all flares prediction studies, the prediction accuracy of weak flares is always higher than that of strong flares. The main reason is that the data is imbalanced. The number of FIs bigger than 10 accounts for just 19% of the total in our data set. There are only 2% of FIs with a total value greater than 100.

Several techniques have been proposed to deal with imbalanced classification tasks (He & Garcia 2009; López et al. 2013). However, the problem of imbalanced domains in the regression task is more complicated. One of the reasons is that the target variable is continuous. In comparison to fixed categorical variables, continuous variables may have an infinite number of possible values. In addition, the definition of more or less relevant values of the target is ambiguous.

In this study, we employ a preprocessing approach called SMOGN (Branco et al. 2017) to deal with imbalanced regression. This technique is effective for skewed distribution-affected machine-learning regression models. The SMOGN algorithm, in particular, when combined with spline regression, can achieve excellent performance (Branco et al. 2017). Essentially, it is a combination of random undersampling (Torgo et al. 2013, 2015) with two oversampling techniques: SMOTER (Torgo et al. 2013) and introduction of Gaussian noise (Branco et al. 2016). By combining random undersampling and oversampling, it is possible to achieve a more balanced distribution of minority and majority cases while reducing their bias.

First of all, the SMOGN algorithm begins by separating the training set into two parts: $P_N$ for majority (normal or less important) part, and $P_R$ for rare but important part. $t_E$ denotes the user-defined relevance threshold for defining the sets $P_N$ and $P_R$. $u\%$ and $o\%$ represent the percentage of undersampling and oversampling, respectively.

Random undersampling is a technique in which observations from the majority part $P_N$ are randomly removed. This sample is then combined with the observations from the minority part to create the final training set for the selected learning algorithm.

The oversampling will use either SMOTER or the introduction of Gaussian noise strategy to generate new cases for the new "more balanced" training set. One of the most widely used approach to synthesizing new examples for classification task is called the Synthetic Minority Over-sampling Technique (SMOTE; Torgo et al. 2013). The SMOTER is a variant of SMOTE for addressing regression.

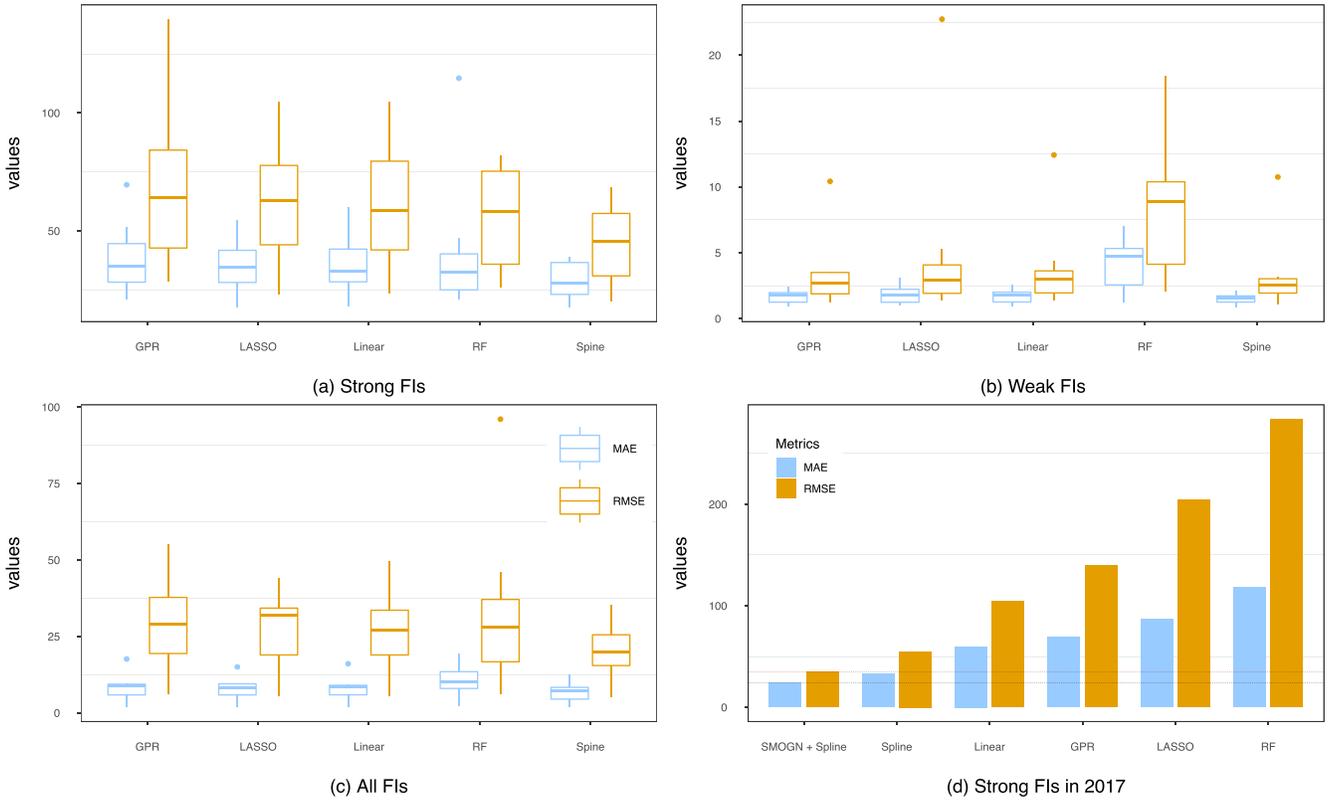

**Figure 6.** Boxplots of the RMSE (orange) and MAE (blue) of the predictions on the (a) strong FIs (FI > 10), (b) weak FIs, and (c) all FIs. (d) The RMSE and MAE of the predictions on the strong FIs in 2017. The vertical dashed lines highlight the minimum values of the RMSE and MAE.

According to the distance between the seed example and all the remaining cases in $P_R$ under consideration, the user is required to set a threshold for determining whether the $k$-nearest neighbor is at a safe or unsafe distance and thus choose between SMOTER and the introduction of Gaussian noise. SMOTER is used if the selected neighbor is deemed "safe." Otherwise, it is preferable to generate a new case by introducing Gaussian noise.

Similarly, we select data from 2010 to 2016 as the training set and data in 2017 as the testing set. The set of FIs greater than 10 was designated as the minority part $P_R$, and the remainder as the majority part $P_N$. In order to confirm the robustness of synthetic data, we produced data sets by applying SMOGN algorithm to original data with fixed sampling ratios of minority to majority (i.e., 30:70, 35:75, 40:60, and 50:50), respectively (Hasanin et al. 2019). One hundred data sets were generated for each resampling ratio. Here, to avoid the issues caused by high dimensionality of the data, only 11 SHARP parameters obtained from Section 2.3.2 are used. Meanwhile, we examined the boxplots of each sample, confirming that the synthetic data reflects the physical significance of solar flare. The SMOGN Python library (Branco et al. 2017) utilized here is available at https://github.com/nickkunz/smogn. Based on our experiments, SMOGN is safe to use when parameter $k$ (the number of neighbors to consider for interpolation used in oversampling) is set to 6 or 7. The results in Figure 5 are obtained by implementing five machine-learning algorithms (i.e., spline regression, RF, linear regression, GPR, and LASSO) on these synthetic data sets. The combination of spline regression and SMOGN shows an optimal performance with $\text{RMSE}_{\min} = 10.53448$, which is evidently smaller than the RMSE = 18.82912 by spline regression alone. Additionally, as shown in Figure 6(d), SMOGN also enhances the performance in forecasting a large FI.

**Table 3**
Nine Feature Rankings Algorithms, and Respective R Package Used

| Learner | Parameter Variants |
| --- | --- |
| Wald Test | rms (Harrell 2021) |
| LASSO | glmnet (Simon et al. 2011) |
| Random Forest (RF) | caret (Kuhn 2021) |
| Bagging Multivariate Adaptive Regression Splines (BMARS) | caret (Kuhn 2021) |
| Fisher Score | Rdimtools (You 2021) |
| DALEX | DALEX (Biecek 2018) |
| Boruta | Boruta (Kursa & Rudnicki 2010) |
| Stepwise Forward and Backward Selection | stats (R Core Team 2021) |
| Relative Importance from Linear Regression (RILR) | relaimpo (Gromping 2006) |

## 4. Magnetic Parameter Ranking

As mentioned earlier in Section 2.3.1, 18 SHARP parameters are evaluated to be statistically significant for the FI prediction. However, we cannot effectively rank the importance of these 18 SHARP parameters because almost all of them have a $p$-value of zero. In this section, we sort the importance of these 18 parameters by the Borda count method calculated from the ranks that are rendered by nine different machine-learning methods (see Table 3: Wald test, LASSO, RF, bagging MARS, Fisher score, Boruta, DALEX, stepwise selection, and relative importance from linear regression). Inspired from voting rules in electoral systems, the idea of the Borda count method here is to combine independent feature ranking results into a more reliable feature ranking. For each ranking, the lowest-ranked feature gets 1 point, the second-lowest-ranked feature gets 2 points, and so on, until the highest-ranked feature gets points equal to the number of features in all "votes," and the total score for each feature is called the Borda count. The Borda count is sometimes referred to as a consensus-based voting method due to its ability to choose a more widely acceptable choice above the majority-supported one; see Green-Armytage et al. (2016).

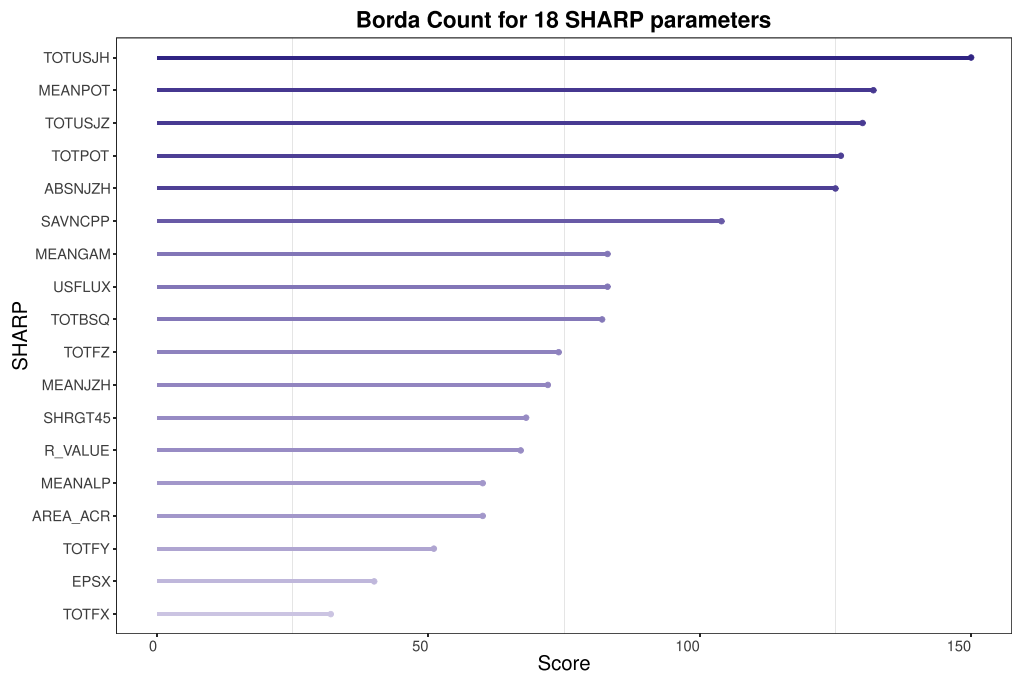

**Figure 7.** The corresponding Borda count for each SHARP parameter based on the nine feature ranking results.

Table 4
Nine Feature Rankings for 18 SHARPS

| SHARP | Wald Test | LASSO | RF | BMARS | Fisher Score | DALEX | Boruta | Stepwise | RILR |
|---|---|---|---|---|---|---|---|---|---|
| TOTUSJH | 1 | 2 | 4 | 3 | 3 | 3 | 3 | 1 | 1 |
| TOTBSQ | 17 | 14 | 8 | 5 | 2 | 7 | 13 | 16 | 7 |
| TOTPOT | 2 | 9 | 6 | 2 | 5 | 6 | 9 | 4 | 2 |
| TOTUSJZ | 10 | 5 | 1 | 7 | 1 | 2 | 1 | 10 | 4 |
| ABSNJZH | 8 | 6 | 2 | 6 | 4 | 4 | 4 | 6 | 6 |
| SAVNCPP | 3 | 10 | 10 | 8 | 7 | 10 | 12 | 2 | 5 |
| USFLUX | 16 | 7 | 5 | 14 | 8 | 5 | 7 | 17 | 9 |
| AREA_ACR | 14 | 15 | 9 | 12 | 11 | 11 | 16 | 13 | 10 |
| MEANPOT | 6 | 4 | 3 | 1 | 10 | 1 | 2 | 9 | 3 |
| R_VALUE | 9 | 8 | 15 | 10 | 15 | 13 | 10 | 12 | 12 |
| SHRGT45 | 12 | 16 | 12 | 11 | 16 | 12 | 6 | 3 | 15 |
| MEANGAM | 4 | 1 | 14 | 9 | 17 | 14 | 11 | 5 | 13 |
| MEANJZH | 11 | 17 | 11 | 4 | 14 | 9 | 5 | 14 | 14 |
| MEANALP | 18 | 18 | 7 | 16 | 6 | 8 | 15 | 15 | 8 |
| TOTFX | 15 | 13 | 18 | 17 | 13 | 18 | 17 | 11 | 17 |
| TOTFY | 7 | 11 | 16 | 15 | 12 | 17 | 18 | 8 | 16 |
| TOTFZ | 5 | 12 | 17 | 13 | 9 | 15 | 8 | 7 | 11 |
| EPSX | 13 | 3 | 13 | 18 | 18 | 16 | 14 | 18 | 18 |

Table 4 and Figure 7 display the nine ranking results and final scores obtained from the Borda count method, respectively. The top 10 SHARP parameters are TOTUSJH, MEANPOT, TOTUSJZ, TOTPOT, ABSNJZH, SAVNCPP, USFLUX, MEANGAM, TOTBSQ, and TOTFZ. We compared our results with the results of similar previous studies (e.g., Bobra & Couvidat 2015; Liu et al. 2017, 2019a; Chen et al. 2019; Wang et al. 2019; Yeolekar et al. 2021). Due to inconsistencies in data and response variables, there are ranking differences in the top 10 SHARP features identified by different strategies. Notably, by comparing our customized rankings, feature subset selection (Yeolekar et al. 2021), and the Fisher score (Bobra & Couvidat 2015; Liu et al. 2017), we find that 9 of these 10 best-performing parameters are identical, indicating the robustness of our ranking results. In the rankings of Chen et al. (2019) and Liu et al. (2019a), the top 10 features change slightly, but the 10 best-performing features in our ranking continue to rank well in their rankings.

The top overall ranking here is TOTUSJH (total unsigned current helicity) with a Borda count of

$$\text{TOTUSJH: } 18(3) + 17(1) + 16(4) + 15(1) = 150. \quad (14)$$

TOTUSJH is calculated as $\sum |\boldsymbol{B}_z \cdot \boldsymbol{J}_z|$, where $\boldsymbol{B}_z$ and $\boldsymbol{J}_z$ are the normal component of magnetic field and electric current density only measured at the photospheric level of the Sun. TOTUSJH, as an approximation to the volume integral of $\boldsymbol{B} \cdot \boldsymbol{J}$, which characterizes the local linkage of elementary currents (Démoulin 2007), is an important proxy measure of an AR's magnetic nonpotentiality (i.e., the degree of deviation from the potential state). It also outperforms other parameters in many previous studies of flare forecasting (e.g., Bobra & Couvidat 2015; Liu et al. 2017, 2019a; Chen et al. 2019; Yeolekar et al. 2021).

The next top ranking parameters are MEANPOT (mean photospheric magnetic free energy; $\frac{1}{N}\sum (\boldsymbol{B}^{\text{Obs}} - \boldsymbol{B}^{\text{Pot}})^2$), TOTUSJZ (total unsigned vertical current; $\sum |J_z| dA$), TOTPOT (total photospheric magnetic free energy density; $\sum (\boldsymbol{B}^{\text{Obs}} - \boldsymbol{B}^{\text{Pot}})^2 dA$), and ABSNJZH (absolute value of the net current helicity; $|\sum \boldsymbol{B}_z \cdot \boldsymbol{J}_z|$). The top five parameters, with the exception of MEANPOT, which is an average-type (intensive) parameter, are all extensive parameters that scale with AR's size. Of these top five parameters, two relate to current helicity (TOTUSJH and ABSNJZH), two to magnetic free energy (MEANPOT and TOTPOT), and the other one measures vertical electric current (TOTUSJZ). Despite their different physical definitions, these parameters are all taken as an important proxy for the magnetic nonpotentiality of the AR. Their high ranking in predicting FI indicates that, compared to other magnetic parameters, current helicity, magnetic free energy, and electric current are key factors in flare forecasting.

## 5. Conclusions

In this paper, we generate flare indices (1 day period) for each AR from 2010 May to 2017 December. Twenty-five SHARP parameters, which produced by the SDO/HMI team, are selected for our data samples for FI prediction. The Yeo–Johnson transformation supports spline regression models by normalizing data. Furthermore, certain correlations between SHARP and the FI emerge after the processing by this transformation. Then, we test statistical significance of each SHARP parameter and the existence of a linear relationship between the FI and SHARPs. Based on the results of hypothesis testing, we establish a preliminary model. Following that, according to the correlation matrix of 25 SHARP parameters, an exhaustive sieving procedure is utilized to eliminate some highly correlated features from the model. After feature selection, we adopt multivariate spline regression algorithm to predict the FI and compare its performance to that of four popular machine-learning methods (i.e., linear regression, LASSO, RF, and GPR). Here, the data sample from 2010 to 2016 is used for training, while the data sample from 2017 is used for testing. Finally, in order to improve the accuracy of strong flare index prediction, SMOGN, an advanced resampling method, is introduced to solve the problem of data imbalance.

Along with the FI prediction, we rank the importance of the SHARP parameter and derive a physical interpretation based on this ranking. The main results of this paper are summarized as follows.

1. Certain correlations between some SHARP parameters and the FI emerge after the processing by the Yeo–Johnson transformation. The linearity test in Section 2.3.1 quantifies this fact through the correlation coefficient value. Nine SHARP parameters (TOTUSJH, TOTBSQ, TOTPOT, TOTUSJZ, ABSNJZH, SAVNCPP, USFLUX, MEANPOT, MEANALP) present a large correlation coefficient value, illustrating the existence of linear relationship with the FI. This is the first time ever that the relationship between SHARPs and the FI has been established.

2. 18 of the 25 SHARP parameters are demonstrated statistically significance in FI prediction by hypothesis testing in Section 2.3.1. Meanwhile, TOTUSJH is shown to be the most important SHARP parameter in Section 2.3. It also outperforms other parameters in many previous studies of flare forecasting. MEANPOT, TOTUSJZ, TOTPOT, and ABSNJZH are the following four ranking factors. This finding suggests that current helicity, magnetic free energy, and electric current are the physical controlling factor of the flare productivity of ARs. Furthermore, certain parameters can still be omitted from the model due to the existing of highly correlation. Eventually, we select 11 parameters for the FI prediction model.

3. Spline regression is demonstrated to be the most effective method for FI prediction among the five tested algorithms (RF, LASSO, GPR, linear regression, spine regression), with an average of RMSE and MAE of 18.9858 and 4.1932, respectively. On this premise, the combination of the SMOGN technique and spline regression considerably improves the accuracy of FI prediction in our study, particularly for a large FI. Notably, there are several considerations to keep in mind when generating data, including the fact that the data's reliability, ratio, etc. will have a substantial influence on the results of the prediction.

On the basis of our findings, we conclude that utilizing SHARP parameters and the spline regression algorithm is a valid method for FI forecasting. It will be fascinating to identify additional effective parameters (such as time-dependent variables) that may be used to further improve the FI prediction.

We thank the team of SDO/HMI for producing vector magnetic field data products. We appreciate deeply the comments and suggestions made by the referee, which improved the paper significantly. The work is supported by US NSF under grants AGS-1927578, AGS-1954737, AGS-2149748, AST-2204384 and AGS-2228996; NASA grants 80NSSC21K1671 and 80NSSC21K0003.

### ORCID iDs

Hewei Zhang ⓘ https://orcid.org/0000-0002-1858-5954
Qin Li ⓘ https://orcid.org/0000-0002-3669-1830
Yanxing Yang ⓘ https://orcid.org/0000-0003-2521-6441
Ju Jing ⓘ https://orcid.org/0000-0002-8179-3625
Jason T. L. Wang ⓘ https://orcid.org/0000-0002-2486-1097
Haimin Wang ⓘ https://orcid.org/0000-0002-5233-565X
Zuofeng Shang ⓘ https://orcid.org/0000-0003-1125-2302